\documentclass[twocolumn,amsmath,amssymb,pre,superscriptaddress,floatfix]{revtex4}

\usepackage{graphicx}
\usepackage{dcolumn}
\usepackage{bm}
\usepackage{amssymb}
\usepackage{multirow}
\usepackage{xcolor}

\newcommand{\1}{\begin{equation}}
\newcommand{\2}{\end{equation}}
\newcommand{\ea}{\begin{eqnarray}} 
\newcommand{\ee}{\end{eqnarray}}

\begin{document}
\title{\fontsize{11pt}{18pt}\selectfont Strategic Spatiotemporal Vaccine Distribution Increases the Survival Rate in an Infectious Disease like Covid-19}

\date{\today}

\author{Jens Grauer}
\affiliation{Institut f\"ur Theoretische Physik II: Weiche Materie, Heinrich-Heine-Universit\"at D\"usseldorf, D-40225 D\"usseldorf, Germany}%
\author{Hartmut L\"owen}
\email[]{hlowen@hhu.de}
\affiliation{Institut f\"ur Theoretische Physik II: Weiche Materie, Heinrich-Heine-Universit\"at D\"usseldorf, D-40225 D\"usseldorf, Germany}%
\author{Benno Liebchen}
\email[]{liebchen@fkp.tu-darmstadt.de}
\affiliation{Institut f\"ur Festk\"orperphysik, Technische Universit\"{a}t Darmstadt, 64289 Darmstadt, Germany}

\begin{abstract}
Covid-19 has caused 
hundred of thousands of deaths and an economic damage amounting to trillions of dollars, creating a desire for the rapid development of vaccine. Once available, vaccine is gradually produced, 
evoking the question on how to distribute it best.   
While official vaccination guidelines largely focus on the question to whom vaccines should be provided first (e.g. to risk groups), 
here we propose a strategy for their distribution in time and space, which sequentially prioritizes regions with a high local infection growth rate. 
To demonstrate this strategy, we develop 
a simple statistical model describing the time-evolution of infection patterns and their response to 
vaccination, for infectious diseases like Covid-19. For inhomogeneous infection patterns, locally well-mixed populations 
and basic reproduction numbers $R_0\sim 1.5-4$ the proposed strategy at least halves the number of deaths in our simulations
compared to the 
standard practice of distributing vaccines proportionally to the population density. For $R_0\sim 1$ we still find a significant increase of the survival rate. 
The proposed vaccine distribution strategy can be further tested in detailed modelling works and could excite
discussions on the importance of the spatiotemporal distribution of vaccines for official guidelines.
\end{abstract}
\maketitle

\noindent\textbf{Introduction:}\\ 
The Covid-19 pandemic 2019/2020 \cite{zhou2020,wu2020,li2020early,wang2020abnormal,of2020species}
has led to almost 10 million infections and 500.000 deaths worldwide (June 2020) \cite{JH,dong2020interactive} 
and an unprecedented social and economic cost which comprises a sudden rise of the number of unemployments by more than 20 million in the USA alone, and a damage of trillions of dollars at the stock market and in the worldwide real economy.
This situation challenges politicians to decide on suitable measures and 
researchers to explore their efficiency, based on models allowing to forecast and compare  
the evolution of infectious diseases (like Covid-19) when taking one or the other action. 
\\Available measures to efficiently deal with epidemic outbreaks at low infection numbers include 
a rigorous contact-tracing (e.g. based on ``Corona-Apps'' \cite{ferretti2020quantifying}) and -testing combined with 
quarantine of infected individuals \cite{lee2020interrupting,wilder2020,mizumoto2020estimating,maier2020effective}. 
Strict travel restrictions preventing an infectious disease
from entering disease-free regions 
(or to die out locally \cite{bittihn2020containment}) present an alternative measure \cite{kraemer2020effect,tian2020investigation}, 
whereas travel reductions by less than $\sim 99\%$ \cite{ferguson2006strategies} slow down the spreading of the disease only slightly \cite{ferguson2006strategies,chinazzi2020effect,anzai2020assessing}.
At higher infection numbers 
reducing the contact rate through measures broadly affecting a population's everyday life, such as 
social distancing \cite{glass2006,wilder2020,stein2020covid,anderson2020will,maier2020effective,vrugt2020effects} and lock-down \cite{lau2020positive,maier2020effective},
remains as the only possibility 
to avoid an explosion of infection numbers. 
Unless a population persistently reduces the contact rate to the point where infection numbers decrease
(this requires a contact reduction by $>60\%$ for a basic reproduction number of $R_0=2.5$ \cite{anderson2020will}), at such stages it has
to accept that the majority of its members has to 
endure the disease -- until finally reaching herd immunity \cite{fine2011herd}.
\\The main hope which remains at such stages rests on the rapid discovery and admission of vaccine \cite{lurie2020developing,graham2020rapid} (or antibodies \cite{antibodies}) to accelerate reaching herd immunity.  
However, while every day where an infectious disease like Covid-19 is active may cause 
thousands of additional deaths, even after admission, 
it may take months until sufficient vaccine is available to overcome an infectious disease. 
Therefore it is important to strategically distribute the available vaccines 
such that the number of deaths remains as small as possible. 
Surprisingly, both official vaccination guidelines, e.g. 
for pandemic influenza \cite{usinfl,whoguidelines}, and previous works on vaccine distribution \cite{sah2018optimizing,medlock2009optimizing,tuite2010optimal}, focus on the question to whom vaccine should be mainly provided, e.g. to prioritize individuals by age or disease risk, and leave 
the quest for a suitable spatial and temporal vaccine distribution 
aside. (Other works like \cite{hethcote1973optimal} ask for the optimal vaccine production rate.) This results in the common practice of simply distributing vaccines proportionally to the population density \cite{venkatramanan2019optimizing}. 
\begin{figure*}
\includegraphics[width=0.85\textwidth]{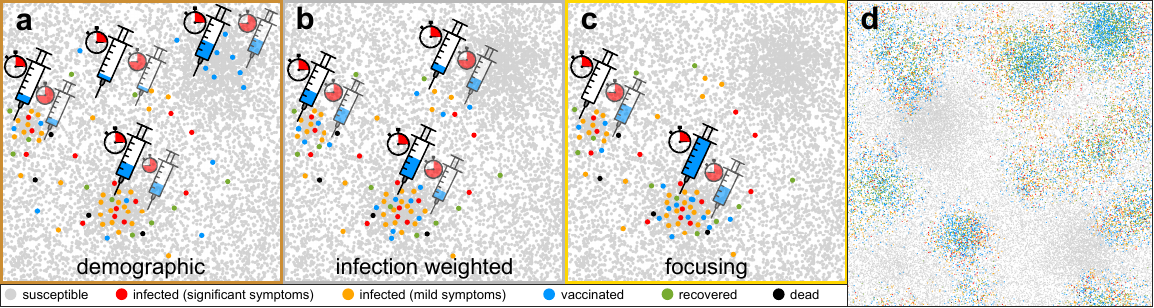}
\caption{Schematic illustration of the proposed spatiotemporal vaccine distribution strategies and of the simulation model. 
a) shows the standard ``demographic strategy'', where 
vaccines (dosage needles) are continuously distributed among all regions (e.g. cities) proportionally to their population density (dots represent groups of individuals).
b) shows the ``infection weighted'' strategy, where vaccines are distributed proportionally to the local infection rates (red and orange dots) and 
c) shows the ''focusing strategy`` where at early times (clocks) only the region with the largest infection rate receives vaccines until the growth rate of a second region catches up and also receives vaccines. 
d) shows a typical simulation snapshot for a city size distribution following Zipf's law taken 42 days after the onset of vaccination when following the focusing strategy. The legend below a)-c) shows the states in our model.}
\label{fig1}
\end{figure*}
\\In the present work we propose an alternative strategy for the spatiotemporal distribution of gradually produced vaccines, 
which hinges on the idea that 
the number of deaths due to a spreading infectious disease is controlled by the pattern of local infection rates, not by population density. 
This strategy, which we call the ''focusing strategy``, 
sequentially prioritizes regions (cities) with the highest growth rates of the local infection numbers (see Fig.~\ref{fig1} and the supplementary movie) and provides, or ''focuses``, all available vaccines to those regions.  
To compare the focusing strategy with the ``demographic'' vaccine distribution practice, 
we develop a simple statistical model describing the time-evolution of an epidemic outbreak (such as Covid-19) and its response to vaccination. 
As our central result, we find that the number of deaths 
resulting from infections occurring after the onset of vaccine production 
is \emph{generally} smaller
when following the focusing strategy rather than the demographic distribution practice.
In fact, for strongly inhomogeneous infection patterns, the focusing strategy reduces the number of deaths by more than a factor of two, 
for a large range of 
basic reproduction numbers $R_0$ and vaccine production rates. The difference is largest for
$R_0\sim 2-3$, 
as might be typical for Covid-19 if no additional measures are in action, but even for $R_0\sim 1$ the focusing strategy significantly increases the survival probability. 
\begin{figure*}
\includegraphics[width=0.85\textwidth]{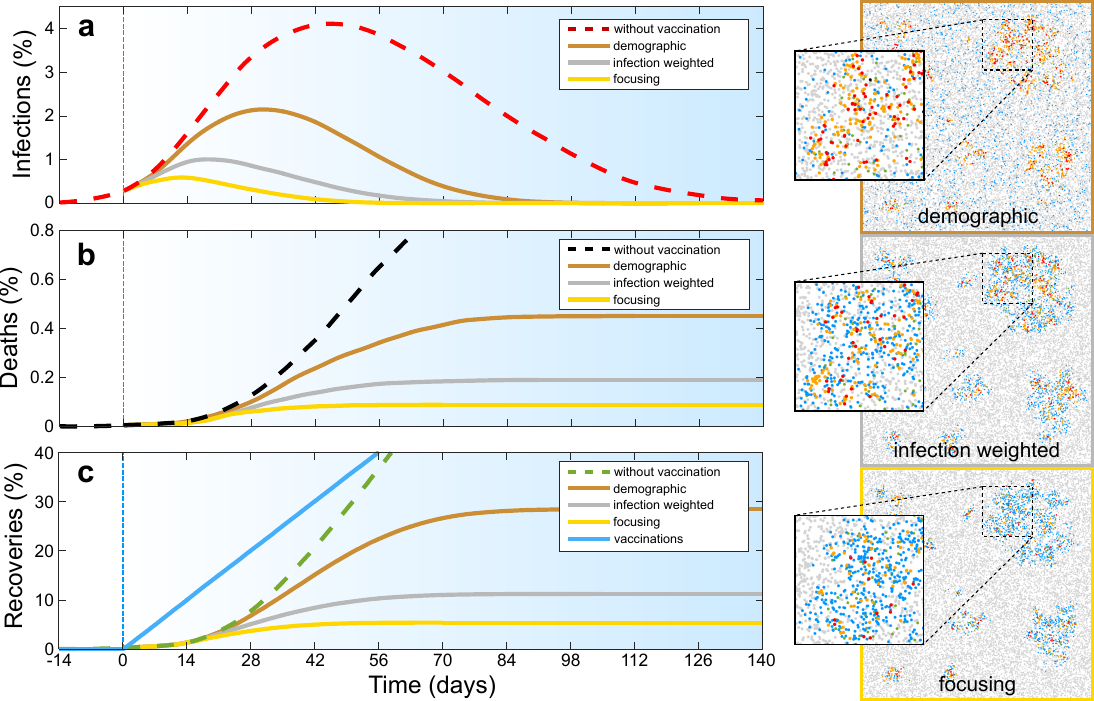} 
\caption{Competition of spatiotemporal vaccine distribution strategies regarding the time evolution of the fraction of infected individuals (a), the fraction of deaths (b), and of recoveries and vaccinations (c). 
Dashed red lines show simulation results without vaccination and bronze, silver (or grey) and gold show results for the demographic vaccine distribution strategy, the infection weighted strategy and the  
focusing strategy respectively. The blue line in panel c) shows the vaccinated fraction of the population 
and vertical blue lines mark the onset of vaccination; the specific time of which is unimportant (see text). 
Panels on the right show simulation snapshots taken 14 days after the onset of vaccine production; insets magnify extracts of these snapshots. 
Parameters: Disease duration 
$t_D=14 \text{days}$; latency time $t_L=t_D/3$, survival probability $s_r=0.965, s_o=0.99$, total vaccination rate $\nu=0.1N/t_D$ and initial reproduction number $R_0=2.5$. 
(The latter is based on $D=10^2 R_c^2/t_D$, $\beta_o=0.3$, $\beta_r=0.1$; see Methods); $L=500R_c$; curves are averaged over 100 random initial ensembles with $N=6000$. 
}
\label{fig2}
\end{figure*}
\section{Model}
To explore the impact of the spatiotemporal vaccine distribution on the disease-evolution in detail, we now introduce a computational model, which is based on Brownian agents and allows deriving a (nonuniform) statistical mean-field model as we will discuss below. Both models are expected to apply to situations where the population is \emph{locally} well-mixed. 
The model describes the dynamics of $N$ agents moving randomly in continuous space in a box of size $L\times L$ with periodic boundary conditions. 
The agents represent groups of individuals and have an internal state variable, 
which is inspired by the SIR model \cite{kermack1927,hethcote2000,marathe2013computational} and its variants 
\cite{pastor2015epidemic,li1995,liu1987dynamical,falco20,adhikari2020inference}. We use colors (see legend in Fig.~\ref{fig1}) to represent the possible states in our simulations, which refer to individuals which are 
``susceptible'' (grey), 
``infected with weak symptoms'' (orange), ``infected with significant symptoms'' (red), 
``recovered'' (green) and ``vaccinated'' (blue). 
Infected agents (orange and red) have an inner clock; they remain symptom free for a latency time $t_L$ and then show 
mild (orange) or significant (red) symptoms for a duration $t_D-t_L$. After an overall disease duration of $t_D$ they 
either recover with a survival probability $s_{o,r}$ (green) or die with probability $1-s_{o,r}$ (black), where the indices refer to agents with mild (orange) and 
significant symptoms (red), respectively.
To model the infection dynamics we describe the spatial motion of an agent with position ${\bf r}_i(t)$ using Brownian dynamics  
$\dot {\bf r}_i(t)=\sqrt{2D}\boldsymbol{\eta}_i(t)$, where $D$ is the diffusion coefficient controlling how fast agents move
and $\boldsymbol{\eta}_i(t)$ represents Gaussian white noise with zero mean and unit variance. 
We assume that all infected agents (orange and red) are infectious, 
both in the latent phase and afterwards (as for Covid-19) and infect a fraction of  
$\beta_o + \beta_r$ of those susceptible agents (grey) which are closer than a distance 
$R_c$; here, indices refer to 
mild (orange) and significant (red) symptoms. 
Agents showing significant symptoms (red) do not move but can infect ``visitors'' if actively approaching them. 
\vskip 0.2cm
To connect the suggested model with standard mean-field descriptions for infectious diseases, we now deduce a continuum model from the Langevin equations describing the agent dynamics. 
The resulting model can be viewed as a generalization of 
standard mean-field models such as the SIR and the SEIR model to inhomogeneous situations and cases where mild and strong infections coexist (as for Covid-19). Let us now consider 
continuous variables (fields) 
representing the local 
mean number density of susceptible agents $S({\bf r},t)$, exposed agents $E({\bf r},t)$ (infected but not yet diseased), 
infected agents which are free of symptoms (or have mild symptoms) $F({\bf r},t)$, infected agents with symptoms $I({\bf r},t)$, recovered (immune) agents $R ({\bf r},t)$ and victims $V({\bf r},t)$. In the absence of social forces (pair attractions, social distancing), the following equations follow by translating Langevin equations to Smoluchowski equations \cite{risken} and coupling them via
suitable reaction terms: 
\begin{eqnarray*}
\dot S({\bf r},t) &=& - \beta' (E+F+I) S/\rho_0 + D \nabla^2 S  - \nabla \cdot (S {\bf f}) - \nu' \\
\dot E({\bf r},t) &=& \beta' (E+F+I) S/\rho_0 - \alpha E + D \nabla^2 E - \nabla \cdot (E {\bf f}) \\
\dot F ({\bf r},t) &=& \alpha r E - \delta F + D\nabla^2 F - \nabla \cdot (F {\bf f}) \\
\dot I({\bf r},t) &=& \alpha (1-r) E - \delta I - \nabla \cdot (I {\bf f}) \\
\dot R ({\bf r},t) &=& \delta (s_o F+ s_r I) + D \nabla^2 R  - \nabla \cdot (R {\bf f}) + \nu' \\
\dot V({\bf r},t) &=& \delta(1-s_r) I + \delta(1-s_o) F
\label{model} 
\end{eqnarray*}
Note here that the exposed state explicitly shows up as a dynamical variable at the continuum level, but only implicitly in our agent-based simulations where infected agents have an inner clock and are in the latent phase before showing (mild) symptoms. 
In the above equations, $\beta'$ is the infection rate, i.e. $1/\beta'$ is the mean time between infectious contacts; $\alpha=1/t_L$ is the rate
to switch from the exposed (latent) state to the infected state, $\delta = 1/(t_D-t_L)$ 
is the recovery rate and $\nu'({\bf r},t)$ is the spatiotemporal vaccination rate which is linked to the constant total vaccination rate 
in the agent-based model via $\nu=\int {\rm d}{\bf r} \; \nu'({\bf r},t)$.
The number $r$ is the ratio of infections proceeding symptom free (or with mild symptoms) and $\rho_0=N/L^2$ is the mean agent density.
Finally, $D$ is the diffusion coefficient and ${\bf f}({\bf r}) = -\nabla_{{\bf r}} U/\gamma$ is the reduced force due to the external potential which we use to create a density profile mimicking a typical city size distribution.
The overall density converges to a Boltzmann distribution 
$S+E+F+I+R+V = N {\rm exp}[-U({\bf r})/(kT)]/ \int {\rm exp}[-U({\bf r})/(kT)] {\rm d} {\bf r}$, yielding the 
conservation law 
$\int (S+E+F+I+R+V)\; {\rm d} {\bf r} =N$ which can be viewed as an expression of the conservation of the overall number density (or the number of agents) in the coarse of the dynamics. 
\\Numerically solving this model by using finite difference simulations now 
allows us to further test the spatiotemporal vaccination strategies. In our simulations we start with the initial state 
$E=F=R=V=0$ and $S=1-\epsilon$, $I=\epsilon$ where 
$\epsilon({\bf r},t)$ represents a small perturbation of the unstable steady state (e.g. $E=F=I=R=V=0, S=1$ for $U=0$), which represents the population before the emergence of the disease. 
The results of these simulations confirm that the spatiotemporal distribution of continously distributed vaccines plays an important role; also here, the infection-weighted strategy and the focusing strategy strongly increase the number of survivors as compared to the demographic distribution. 

\section{Results}
We now perform numerical simulations of both the proposed agent based model and the statistical mean-field model which both lead to consistent results. 
For the agent based model we perform Brownian dynamics simulations \cite{erban2014,romanczuk2012active,volpe2013simulation,stenhammar2013continuum,winkler2015virial,martin2018collective}
starting with $2 \times 10^{-3} N$ randomly distributed initial infections and an initial reproduction number $R_0=2.5$ such that infection numbers exponentially increase over time.   
Let us assume that vaccine production starts after some initial transient and then allows to transfer 
$\nu$ individuals per day from the susceptible to the immune state.   
(Note that the duration of the initial transient is unimportant in our simulations, if 
vaccination starts long before herd immunity is reached.)
Now considering the time-evolution of the percentage of infected, dead and recovered individuals of a given population, 
and distributing the available vaccines 
proportionally to the population density (bronze curves in Fig.~\ref{fig2}), we observe an 
infection maximum (panel a) about 30 days (two infection cycles) after the onset of vaccine production, i.e. when about 22\% of the population have received vaccines and 2\% of the population is infected. 
When distributing the available vaccines proportionally to the local infection rate (``infection weighted strategy'')
instead, notably, the infection 
maximum occurs an entire infection cycle earlier (silver curve in panel a). Here the infection number peaks 
when only 11\% of the population has received vaccines and only 1\% is infected. 
However, the infection weighted strategy is not optimal but can be further improved by exclusively providing 
all available vaccines to the region (e.g. a city) with the highest infection rate
(``focusing strategy''). This means that initially only a single region receives vaccines until the infection rate of 
a second region catches up and both regions simultaneously receive vaccines, until a third region catches up and so on. Following this ``focusing strategy'' the infection peak further shifts to earlier times (golden curve in panel a)
and occurs when only 0.6\% of the population is infected. 
Importantly, the resulting fraction of deaths reduces by more than a factor of two when following the infection weighted strategy (silver) rather than the demographic strategy (bronze). It almost halves again when 
following the focusing strategy instead (gold). 
This shows that the precise spatial and temporal order of vaccine donation controls the number of survivors from an infectious disease. 
\\We now complement these results by numerical solutions of the statistical mean-field model equations by finite-difference simulations. As in our particle based simulations we find that the focusing strategy is generally better than the 
infection-weighted strategy and the demographic vaccine distribution strategy. The results of the agent-based simulations and the 
continuum simulations show a close quantitative agreement (not shown for the uniform system; see Fig.~\ref{fig4} for an exemplaric quantitative comparison in the presence of 'cities'.).

\begin{figure}
\includegraphics[width=0.48\textwidth]{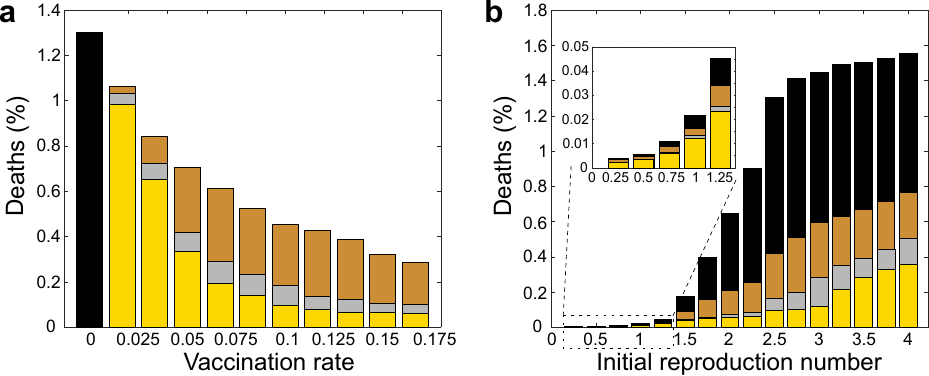} 
\caption{Fraction of deaths as a function of the vaccine production rate (left) and the initial basic reproduction number (right) for the demographic strategy (bronze), the infection-weighted strategy (silver) 
and the focusing strategy (gold). Results without vaccination (black) are shown for comparison.
The results are based on the agent-based model; the statistical mean-field equations lead to very similar graphs. 
Parameters are shown in the key; remaining ones are as in Fig.~\ref{fig2}.
}
\label{fig3}
\end{figure}

To systematically explore the robustness of these findings 
we now repeat our simulations for different vaccine production rates and initial reproduction numbers.
Fig. \ref{fig3} shows that the resulting fraction of deaths, counted once the disease is gone, is generally highest for the demographic strategy (bronze)
and lowest for the focusing strategy (gold).  
Mathematically, this is because vaccination is most efficient at locations where it maximally reduces the infection growth rate, which holds true independently of the specific parameter regime. 
The differences among the individual strategies is comparatively large if vaccine is produced fast enough to allow vaccinating at least about 1\% of the population per day and at reproduction rates around $R_0\sim 2-3$. The latter value might be sensible for Covid-19. However, even for slower vaccine production or for $R_0\sim 1-2$ (as typical for influenza), several percent of deaths can be avoided in our simulations by strategically distributing the available vaccines in space and time.

To further explore the applicability-regime of the focusing strategy, we now combine it with social distancing rules, which reduce the effective reproduction number to $R_t\sim 1$. We implement the latter as a phenomenological repulsive three-body interaction among the agents (see Methods for details)
which prevents them from aggregating in groups of more than two individuals. 
Also here, the resulting deaths fraction (Fig.~\ref{fig4}a) saturates significantly earlier when following the focusing strategy (gold) rather than the demographic strategy (bronze). The difference in deaths numbers among the three different vaccination strategies is almost identical 
to our corresponding results at $R_0\sim 1$ but without social distancing (Fig. 3b). 

Finally, we explore a possible impact of a nonuniform population distribution (city structure) on the proposed vaccination strategies. We create a population with a spatial density distribution following   
Zipf's law which closely describes the city size distribution in most countries \cite{gabaix1999zipf} as $\tilde P_c(s>S) \propto 1/S$,  where $\tilde P_c(s)$ is the probability that a city is larger than $S$.
To generate a population featuring a corresponding population distribution, we add an 
external potential $U$ to the equation of motion of the agents (see Methods for details). 
Following statistical mechanics, the resulting population density follows Boltzmann's law 
$P({\bf r}) \propto {\rm exp}[-U({\bf r})/(kT)]$ where $P({\bf r})$ is the probability that an agent is at position ${\bf r}$ and $kT=\gamma D$ is the effective thermal energy of the agents, controlling how often agents leave a ``city'' (minimum of $U$). 
Now matching Boltzmann's distribution with Zipf's law yields a construction rule for $U$ (see Methods) to create a
population pattern featuring a characteristic city-size distribution. 
Our resulting simulations, shown in Fig.~\ref{fig4}b, and in the supplementary movie (for $N=55.000$ agents), demonstrate that the 
focusing strategy and the infection weighted-strategy again halve the number of deaths compared to the demographic strategy. Here, the former two strategies are comparatively close to each other regarding the number of resulting deaths, which indicates that 
in strongly inhomogeneous populations a suitable spatial vaccine distribution rule might be even more 
important than the precise temporal sequence of vaccine donation. 
\\To further test the robustness of these findings, we have performed continuum simulations of our statistical mean-field 
model, which leads to close quantitative agreement with the particle based simulations (Fig.~\ref{fig4}b). Typical snapshots of the infection pattern 56 days after the onset of vaccination are shown in Fig.~\ref{fig5}. These figures show a clear reduction of the infection number in all infection hotspots for the focusing strategy (panel c) as compared to the infection weighted strategy (b) and in particular compared to the demographic vaccine distribution practice (a).

\begin{figure}
\includegraphics[width=0.4\textwidth]{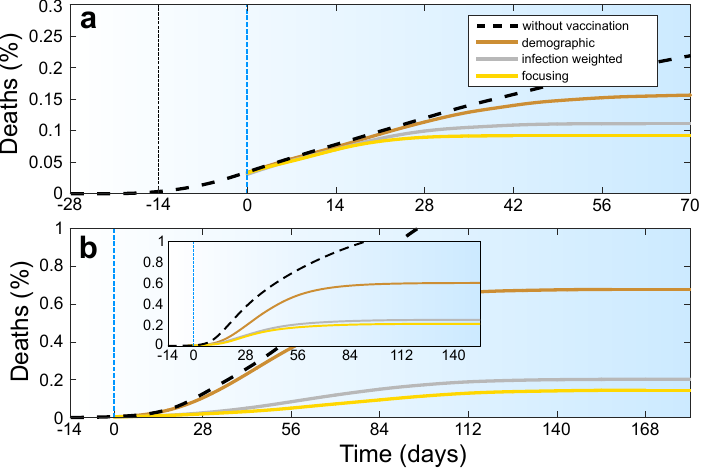} 
\caption{Competition of spatiotemporal vaccination strategies a) in the presence of social distancing which is activated after 14 days (black vertical line) and reduces the reproduction number to $R\approx 1$
and b) for a population density distribution following Zipf's law. 
Colors and parameters are as in Fig.~\ref{fig2} but we have $N=12000$, $L=700$, $R_0=2.7$ (which is based on $D=10^3 R_c^2/t_D$ and $\beta_o=0.05$, $\beta_r=0.017$) and $\nu=0.05N/t_D$. 
Inset: Analogous results for the mean-field model using same parameters as in the agent-based model and   
a 140$\times$140-grid with each grid point corresponding to a spatial area of $5R_c \times 5R_c$.
}
\label{fig4}
\end{figure}

\begin{figure*}
\includegraphics[width=0.96\textwidth]{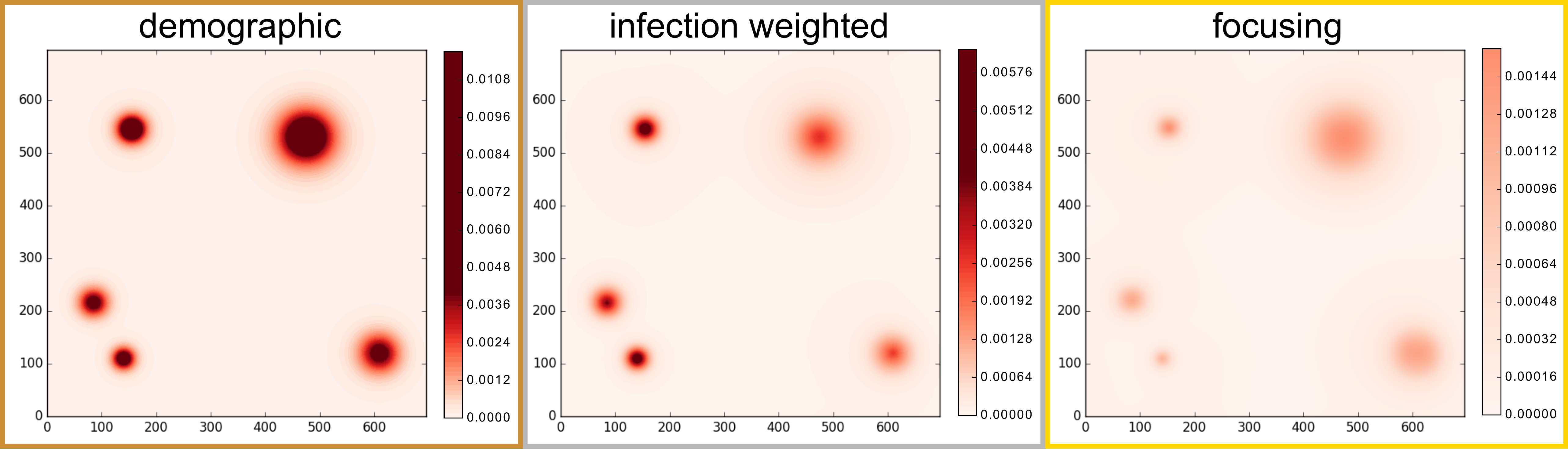} 
\caption{Snapshots of the infection patterns 56 days after the onset of vaccination, based on the statistical mean-field model. Colors show the density of exposed agents $E(\vec r,t)$.
Parameters are as in Fig.~\ref{fig4}b.
}
\label{fig5}
\end{figure*}

\section{Discussion}
Our simulations suggest that a strategic spatiotemporal distribution of gradually produced vaccines 
generically increases the number of survivors in ongoing epidemic disease.
In particular, by sequentially prioritizing spatial regions (cities) 
with the highest local infection growth rates, the proposed ``focusing strategy''
reduces the number of deaths 
by more than a factor of two compared to the standard practice of distributing vaccines demographically. 
Such a strong difference occurs 
for locally well-mixed populations and a  
large range of initial reproduction numbers ($R_0\sim 1.5-4$) and vaccine production rates and even in combination with additional social distancing measures, if the underlying infection pattern is sufficiently inhomgeneous and 
vaccine production starts long before the population reaches herd immunity.  
For $R_0 \sim 1$ we still find a reduction of the fraction of death by up to about $35\%$. 
Our results might excite discussions regarding the importance 
of the spatiotemporal distribution of gradually produced vaccines for official
vaccination guidelines and should be further tested in detailed models e.g. to explore the 
impact of the proposed strategy also in situations where the population is not locally well-mixed and to 
combine the suggested spatiotemporal distribution strategy with the prioritization of certain individuals such as risk groups, 
individuals with a strong social mixing tendency or with jobs of systemic relevance.

\vskip 0.2cm
\paragraph*{Acknowledgements}
We thank Michael E. Cates for helpful comments.

\newpage

\begin{center}
{\fontsize{11pt}{18pt}\selectfont\textbf{Methods}}
\end{center}
\vskip 0.6cm
\noindent\textbf{Simulation details:}
\\To calculate the spatial dynamics of the agents in our model, we solve Langevin equations 
$\dot {\bf r}_i(t)=\sqrt{2D}\boldsymbol{\eta}_i(t)$ with $i=1,..,N$ using Brownian dynamics simulations involving a 
forward Euler time-stepping algorithm and a time-step of $dt=0.0028$ days which amounts to about 4 minutes. 
After each timestep we check for each infected agent (red or orange) which susceptible agents (grey) are closer than $R_c$. We then change the state of the latter agents to an infected state with an infection rate of
${\tilde \beta}_o=3{\tilde \beta}_r=0.0075/dt$ (Figs. 2-4a), corresponding to infections with mild symptoms (orange) and significant symptoms (red), respectively. These rates yield $\beta_o=3\beta_r=0.3$ for the corresponding fractions of 
contacts which lead to infections. 

\vskip 0.6cm
\noindent\textbf{City size structure:}
\\To generate a population density distribution with a structure which is typical for cities, we add an external potential landscape $U({\bf r})$ to the Langevin equations describing the dynamics of the agents, 
i.e. 
$\dot {\bf r}_i(t)=\sqrt{2D}\boldsymbol{\eta}(t) - \nabla_{{\bf r}_i} U({\bf r}_i)/\gamma$. Here $\gamma$ is an effective ``drag'' coefficient determining the strength of the response of the agents to $U$. 
We now create $U$ as a superposition of Gaussians,  
$U(\mathbf{r})=\sum_j a e^{-\frac{(\mathbf{r}-\mathbf{r}_j)^2}{2\sigma_j^2}}$, each of which leads to a population density maximum around $\mathbf{r}_j$, which represents the center of 
city $j$. Here $a$ is the strength (amplitude) of the reduced potential which we choose as $a=D \gamma/2=kT/2$ and $\sigma_j$ defines the radius of city $j$, which we choose randomly 
from a distribution $P(\sigma)=\frac{1}{\sigma} \frac{1}{\ln(R_{\rm max}/R_{\rm min})}$ where $R_{\rm min}=20R_c$ and $R_{\rm max}=80R_c$ are the minimal and the maximal possible ``city radius'' in the simulations underlying Fig.~4b. 
We randomly distribute the city centers ${\bf r}_j$ within the simulation box. 

\vskip 0.6cm
\noindent\textbf{Social distancing:}
\\To effectively model social distancing in a simple way, we phenomenologically add repulsive excluded volume interactions among the agents which prevent that groups of more than two agents form. That is, we choose 
$U=\frac{1}{2} \sum_{k,l \neq k} V_{kl} \nu_{kl}$ where the sums run over all agents and where
$V_{kl}$ represents the Weeks-Chandler-Anderson
interaction potential among agents $k,l$, i.e. 
$V_{kl} = 4\epsilon \left[ (\frac{d}{r_{kl}})^{12} - (\frac{d}{r_{kl}})^6 \right] + \epsilon $
if $r_{kl} \leq 2^{1/6}d$ and $V_{kl}=0$ otherwise.
Here $r_{kl}$ denotes the
distance between agents $k$ and $l$ and $r_{cut}=2^{1/6} d$ represents a cutoff radius beyond which the interaction potential is zero; 
$\epsilon$ controls the strength of the potential and is chosen such that $\epsilon/\gamma=D$.
In our simulations at each timestep we choose $\nu_{kl}=1$ if at least one of the agent $k$ and $l$ has a 
``neighbor'' at a distance closer than $d=3R_c$ and otherwise we choose $\nu_{kl}=0$. 
In addition, we add a weak pair attraction of strength $D/10$ and range $d=3R_c$ to our simulations to support the formation of pairs. 
That way, agents can form pairs but there is a significantly reduced probability that they 
form triplets or larger groups. 
\vskip .6cm

\noindent\textbf{Relation of reproduction number to simulation parameters:}
\\Here we relate the effective reproduction number $R_e(t)$, which is the average number of infections caused by an infected agent at time $t$, with the microscopic parameters in our simulation.
For this purpose, let us first consider the 
area $A(t)$ covered by a Brownian agent with radius $R_c$ and diffusion coefficient $D$ over a time $t$. This area is known as the Wiener sausage  \cite{rataj2009expected} and reads  
\begin{equation}
 A(t) = \pi R_c^2 + \frac{8R_c^2}{\pi} \int_0^{\infty} \frac{1-e^{-\frac{2D y^2 t}{2R_c^2}}}{y^3(J^2_{0}(y)+Y^2_{0}(y))} dy \; ,
\end{equation}
where $J_{0}(y)$ and $Y_{0}(y)$ are the 0-th Bessel functions of the first and second kind.
Now denoting the agent density of susceptible agents with $\rho_S$, the average number of (possibly infectious) contacts during a time $\tau$ is $A(\tau) \rho_S$. 
Thus, if agents are infectious over an overall time of $t_D$ and the fraction of contacts which lead to infections with significant (mild) symptoms is $\beta_r$ ($\beta_o$), we obtain the following 
expression for the (spatially averaged) effective reproduction number $R_e$: 
\begin{equation}
 R_e(t) = A(t_D) \rho_S(t) (\beta_o+\beta_r) \; .
\end{equation}
where $R_e(t=0)=R_0$.
This expression links the reproduction number with the microscopic simulation parameters and reveals that the reproduction number at time $t$ is proportional to the average density of susceptible agents at time $t$. 
\vskip 0.6cm
\noindent\textbf{Supplementary Movie:}\\
The movie shows the time-evolution of the modeled infection pattern for $N=55.000$ agents 
and its response to the proposed spatiotemporal vaccine distribution strategies. 
Parameters are as in Fig. 4b and the population distribution in the movie follows a typical city size structure (Zipf's law).

\bibliography{literature}

\end{document}